\documentclass[final,1p,times]{elsarticle} 
\usepackage{bm}
\usepackage{mathrsfs}
\usepackage{graphicx}
\usepackage{amssymb} 
\usepackage{amsthm} 
\usepackage{lineno}

\journal{Nuclear Physics A} 

\begin{document}

\begin{frontmatter} 

% Your Title - please insert
\title{Viscous Flow in Heavy-Ion Collisions from RHIC to LHC}

%% Single author (and collaboration) - please insert
\author{Chun Shen}
\author{Ulrich Heinz}
\address{Department of Physics, The Ohio State University,
  Columbus, Ohio 43210-1117, USA}

%% Multiple authors
%\author[auth2]{Marcus Junius Brutus}
%\address[auth1]{Somewhere, Rome}
%\address[auth2]{Somewhere else, Rome}

\begin{abstract} 
We present a systematic hydrodynamic study of the evolution of hadron spectra and their azimuthal anisotropy from the lowest collision energy studied at the Relativistic Heavy Ion Collider (RHIC), $\sqrt{s} = 7.7$\,$A$\,GeV, to the highest energy reachable at the Large Hadron Collider (LHC), $\sqrt{s} = 5500$\,$A$\,GeV \cite{Shen:2012vn}. The energy dependence of the flow observables are quantitatively studied for both the Monte-Carlo Glauber and Monte-Carlo Kharzeev-Levin-Nardi (MC-KLN) models. For MC-Glauber model initial conditions with $\eta/s = 0.08$, the differential charged hadron elliptic flow $v_2^\mathrm{ch}(p_T, \sqrt{s})$ is found to exhibit a very broad maximum in the region $39 \le \sqrt{s} \le 2760$\,$A$\,GeV. For MC-KLN initial conditions with $\eta/s = 0.2$, a similar ``saturation'' is not observed up to LHC energies. We emphasize that this ``saturation'' of elliptic flow arises from the interplay between radial flow and elliptic flow which shifts with $\sqrt{s}$ depending on the fluid's viscosity. By generalizing the definition of spatial eccentricity to isothermal hyper-surface, we also calculate $\varepsilon_\mathrm{x}$ on the kinetic freeze-out surface at different collision energies.
\end{abstract} 

\end{frontmatter} % do not change

%% linenumbers are useful for reviewing process
%\linenumbers

\section{Introduction}
The recent Beam Energy Scan (BES) program \cite{Kumar:2011de,Shi:2011ad,QM12BES} at the Relativistic Heavy Ion Collider (RHIC) pursues one of the major goals of heavy-ion collision experiments: to explore the QCD phase diagram and search for the phase boundary between the normal nuclear matter and quark-gluon plasma (QGP). The BES program at RHIC together with Pb+Pb collisions at the Large Hadron Collider (LHC) provide us with a unique opportunity to study systematically the collision energy dependence of a large number of relativistic heavy-ion collision observables.

Here we study the collision energy dependence of charged hadron transverse momentum spectra and elliptic flow coefficients \cite{Shen:2012vn}, using (2+1)-d viscous hydrodynamics coupled with a  modern lattice QCD based equation of state \cite{Huovinen:2009yb, Shen:2010uy}. Our work focuses on qualitative tendencies rather than quantitative comparison with experimental data. It makes simplifying assumptions that are justified at high energies but gradually break down at lower $\sqrt{s}$: longitudinal boost invariance, an equation of state for matter with zero net baryon density, and a purely hydrodynamic approach with constant specific shear viscosity, including for the dilute hadronic rescattering stage which at lower energies occupies an increasing fraction of the fireball's dynamical history and should be described microscopically
\cite{Song:2010mg}. These limitations can be cured in future work; they are expected to modify our conclusions quantitatively but not qualitatively.

\section{Results and discussion}
\noindent
{\bf Evolution of charged hadron multiplicity and total elliptic flow: }
%
%=======================================
\begin{figure*}
  \begin{minipage}{0.79\linewidth}
  \begin{tabular}{cc}
  \includegraphics[width=0.47\linewidth,height=0.35\linewidth]{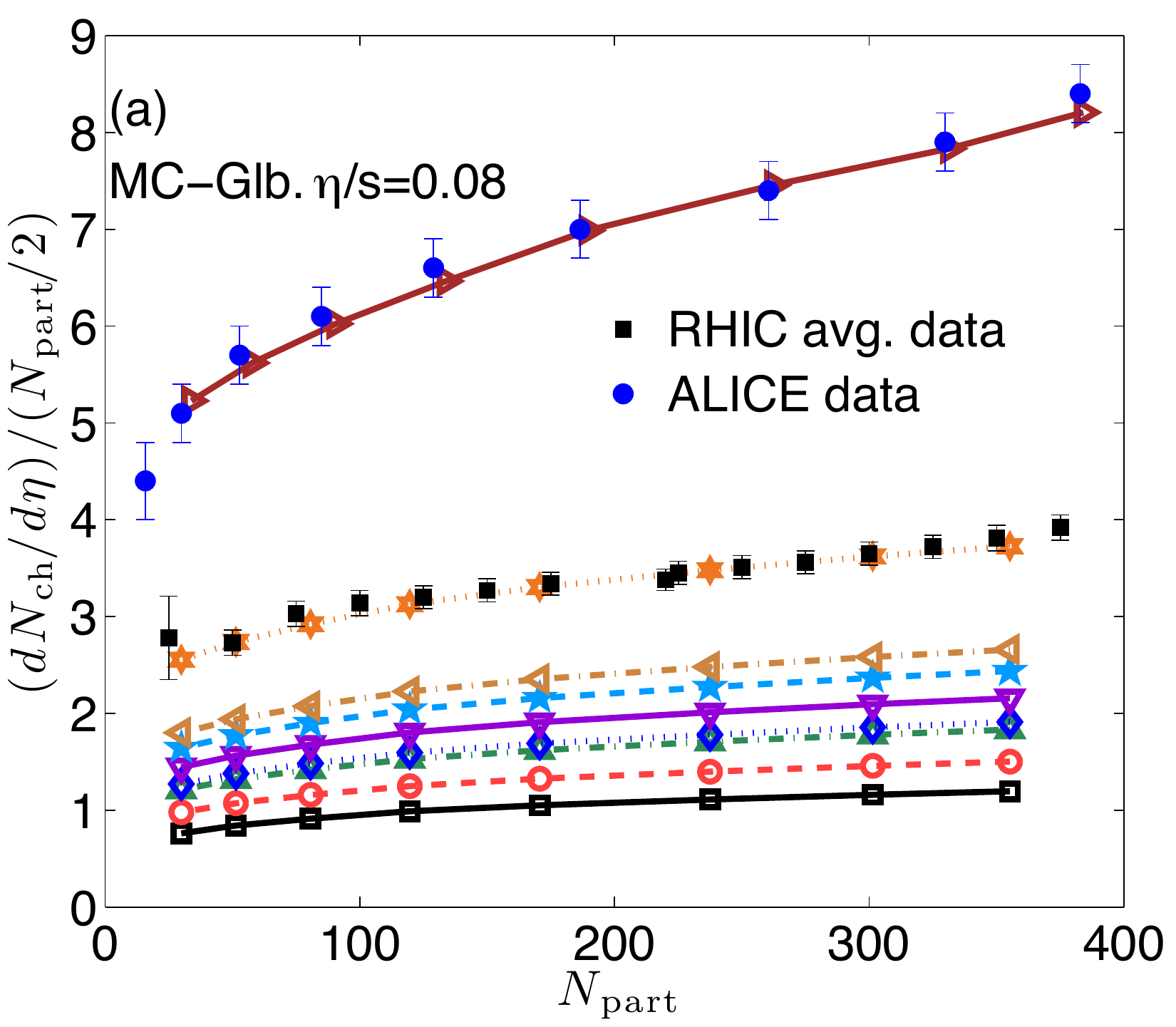} &
  \includegraphics[width=0.47\linewidth,height=0.35\linewidth]{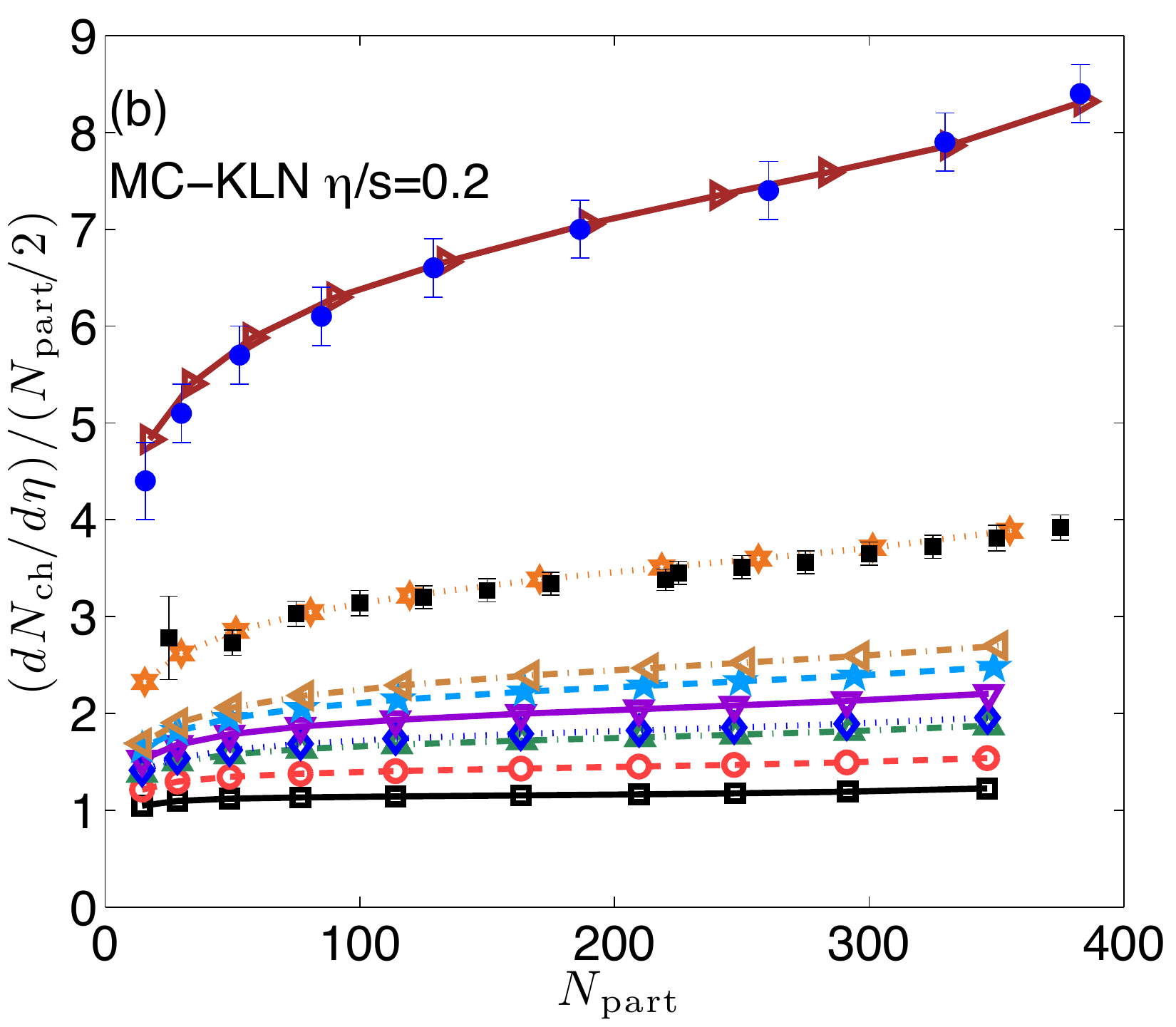} \\
  \includegraphics[width=0.47\linewidth,height=0.35\linewidth]{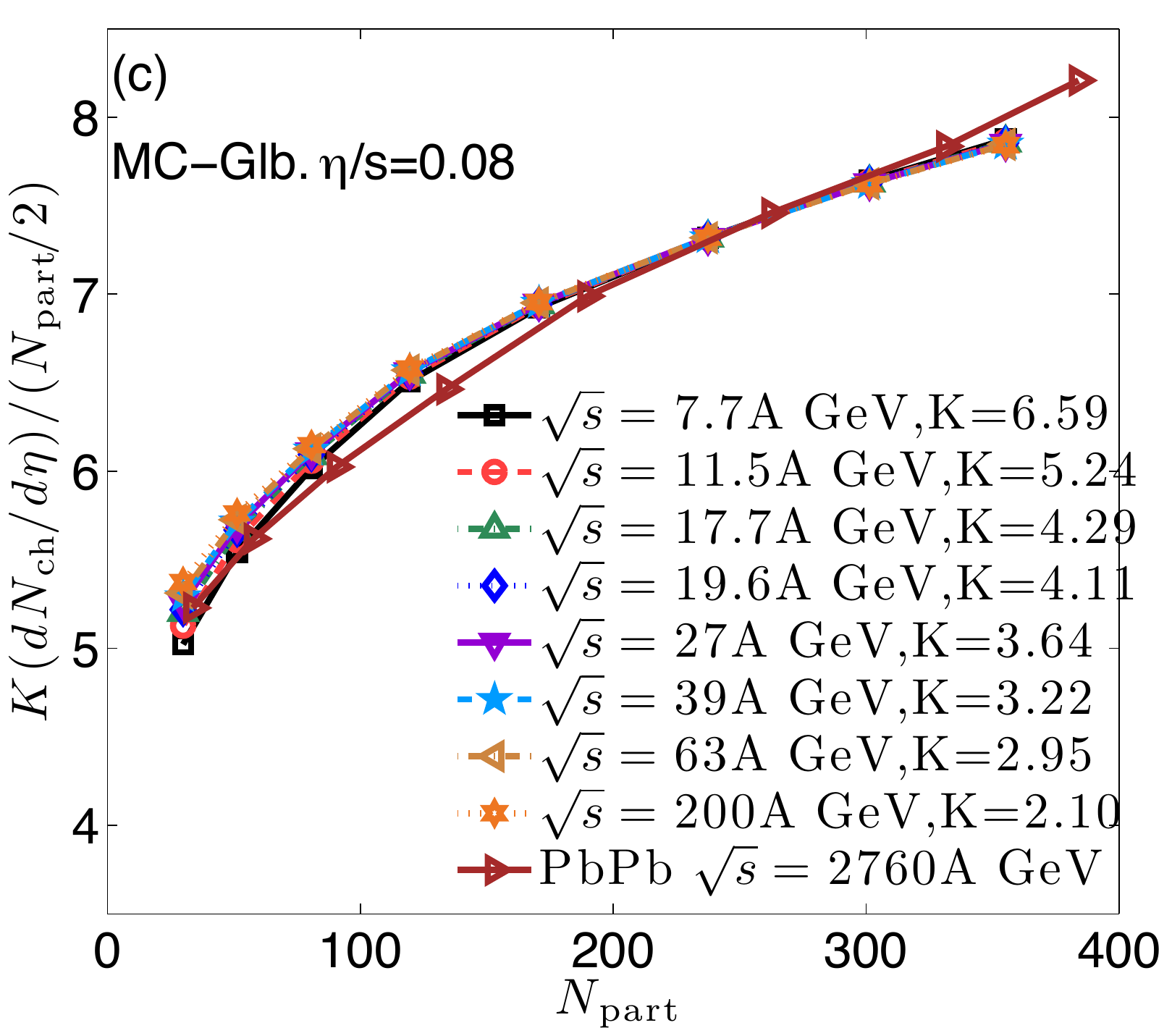} &
  \includegraphics[width=0.47\linewidth,height=0.35\linewidth]{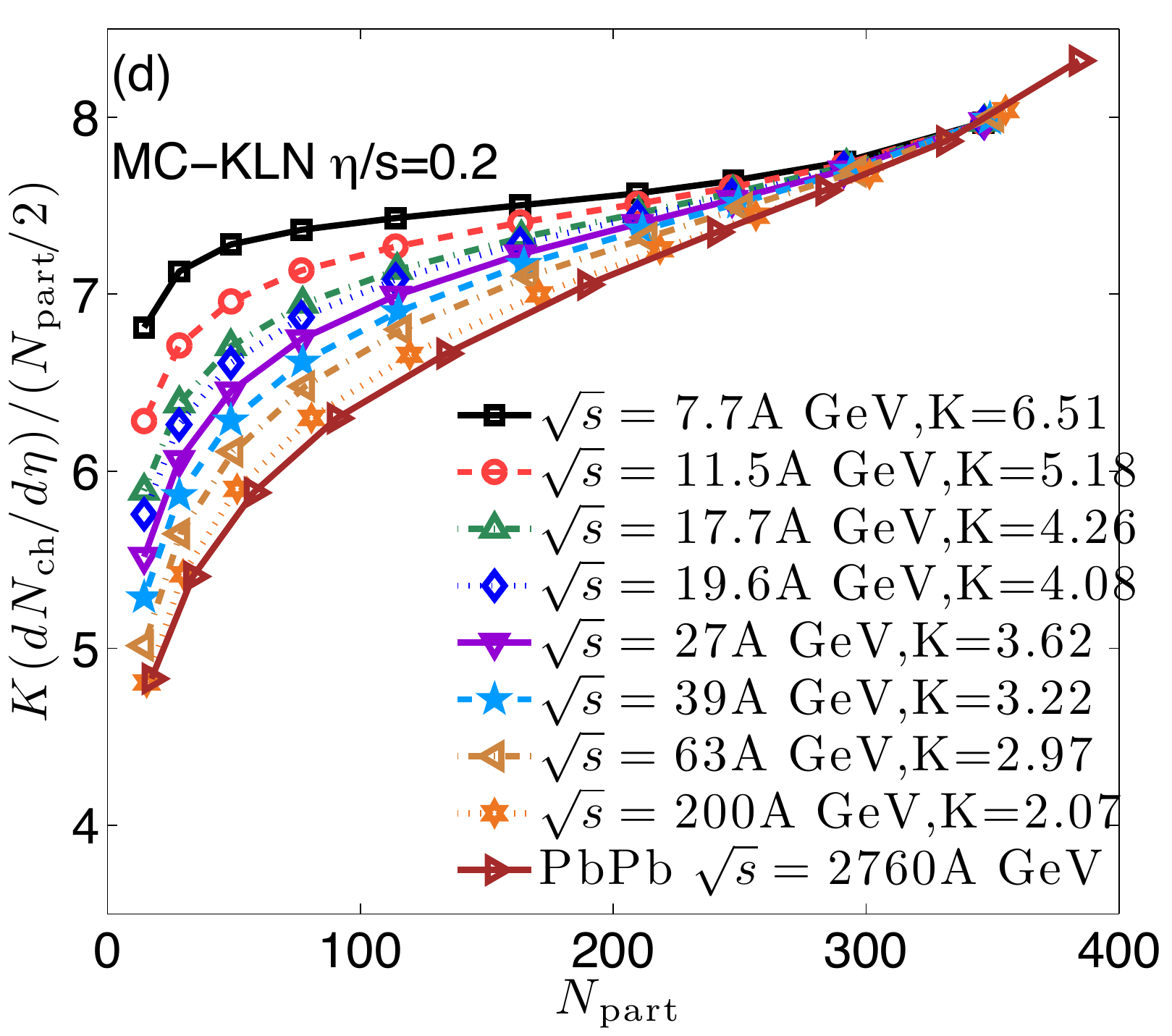}
  \end{tabular}
  \end{minipage}
  \begin{minipage}{0.2\linewidth}
  \caption{{\bf (a):} Centrality dependence of final charged hadron multiplicity per participant nucleon pair as a function of $N_\mathrm{part}$ for MC-Glauber initial conditions, with collision energies varying from $\sqrt{s}{\,=\,}7.7A$\,GeV to $\sqrt{s}{\,=\,} 2760 A$\,GeV. {\bf (c):} Centrality dependence of $\frac{dN_\mathrm{ch}}{d\eta}$ from the lower energy runs in (a) scaled up to the LHC results, for shape comparison. {\bf (b, d):} Same as (a, c) but for MC-KLN initial conditions.}
  \label{fig1}
  \end{minipage}
\end{figure*}
%=======================================
%
In Figs.\,\ref{fig1}(a,b) we show the centrality dependence of the charged hadron multiplicity for both MC-Glauber and MC-KLN models with collision energies from $\sqrt{s}=7.7$ to 2760\,$A$\,GeV.  The curves account for viscous entropy production during the hydrodynamic evolution. At LHC and top RHIC energies our results for both initialization models agree well with the experimental data \cite{Adler:2004zn,Aamodt:2010cz}. Our lower collision energy predictions can be checked against data from the RHIC BES program. In Figs.\,\ref{fig1}c,d we scale the lower energy results by constant factors to align them with the LHC curve in central (0-10\%) collisions, to see how the centrality dependence changes with $\sqrt{s}$. For the MC-Glauber model, the curves fall almost on top of each other. This is because we keep the mixing ratio between the wounded nucleons and binary collisions fixed for the low energy runs at RHIC, and it also reflects the fact that viscous entropy production is small and has little effect on the centrality dependence. For the MC-KLN model, however, the slope of the centrality dependence gets flatter as the collision energy decreases. Only the top RHIC and LHC energy curves approximately fall on top of each other. We found that this tendency originates in the nature of the MC-KLN model itself. Our MC-KLN calculations thus predict a violation of the $\sqrt{s}$-scaling of the centrality dependence of $\frac{dN_\mathrm{ch}}{d\eta}$ at lower collision energies that is not seen with the MC-Glauber initial conditions. This may help to discriminate experimentally between these models.  

%=======================================
\begin{figure*}
  \begin{minipage}{0.79\linewidth}
\begin{tabular}{cc}
  \includegraphics[width=0.47\linewidth]{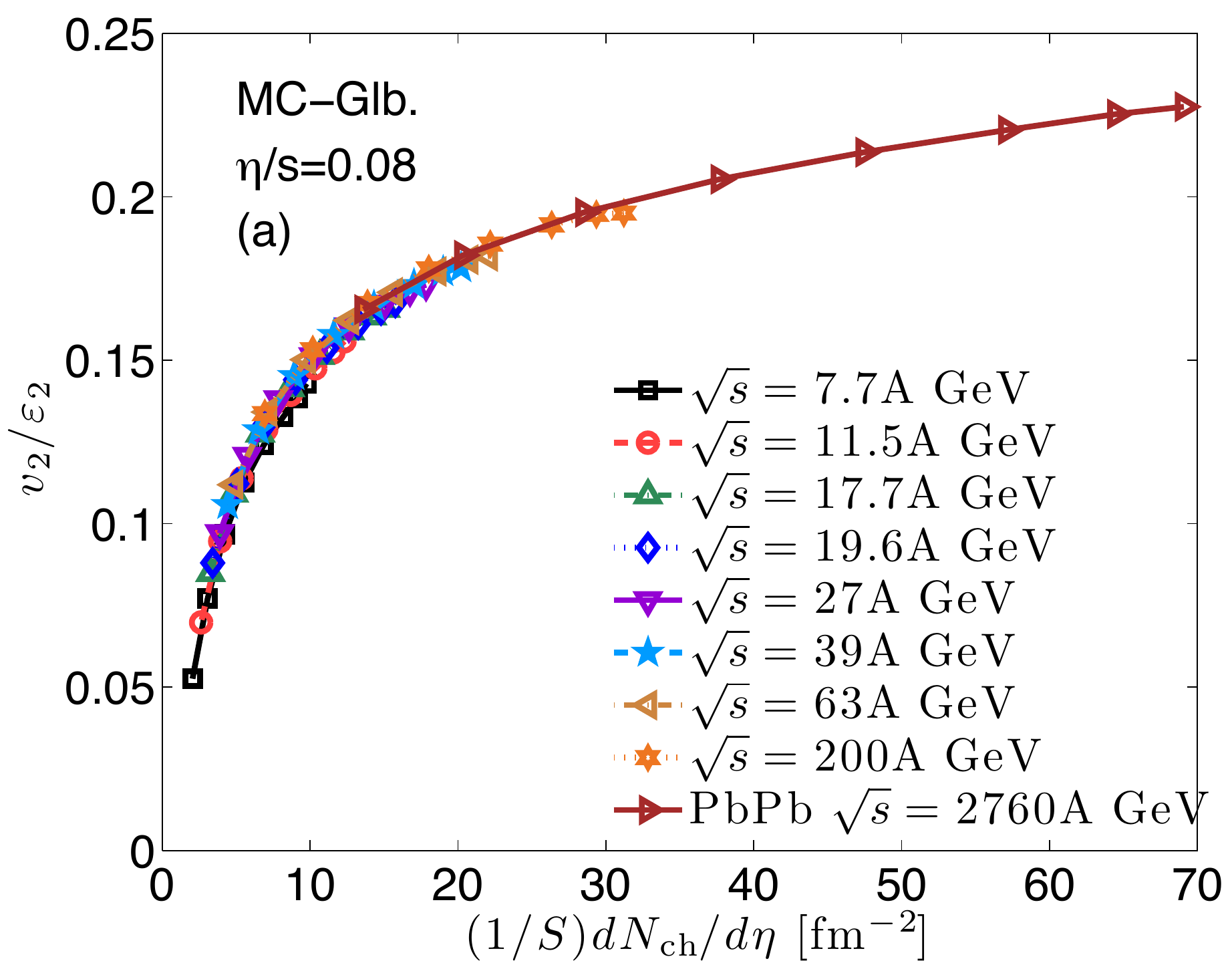} & 
  \includegraphics[width=0.47\linewidth]{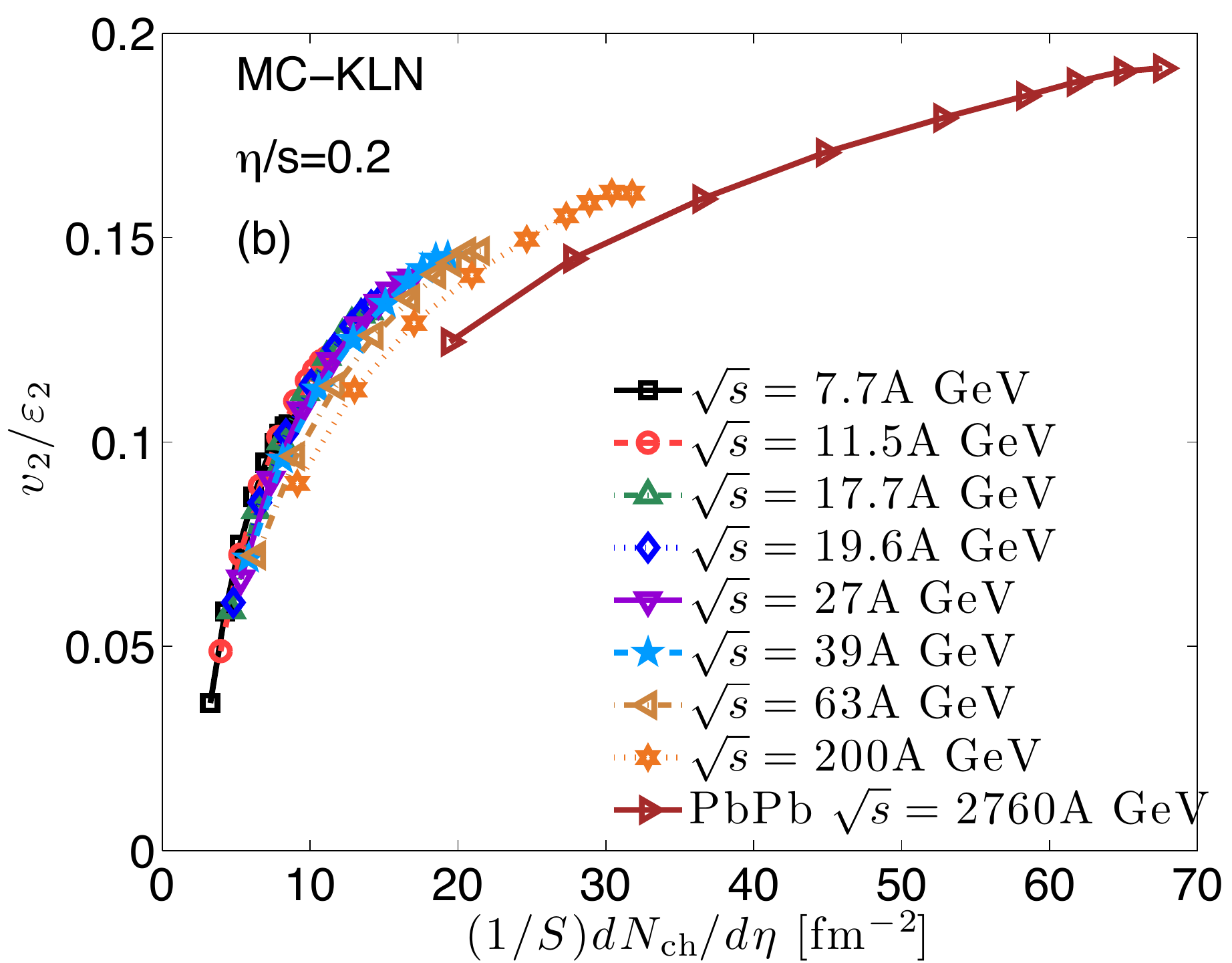}
\end{tabular}
\end{minipage}
\begin{minipage}{0.2\linewidth}
\caption{Eccentricity-scaled $p_T$-integrated $v_2$ plotted as a function of the charged hadron multiplicity density for different collision energies, for MC-Glauber initial conditions with $\eta/s=0.08$ (a) and MC-KLN profiles with $\eta/s=0.2$ (b), respectively. }
\label{fig2}
\end{minipage}
\end{figure*}
%=======================================
% 

In Fig.~\ref{fig2} we explore the scaling of elliptic flow with charged hadron multiplicity density (``multiplicity scaling'') over a wider range of $\sqrt{s}$ than previously studied, for both of the initialization models. For MC-Glauber initial conditions (Fig.~\ref{fig2}a) the ``multiplicity scaling'' curve $v_2/\varepsilon_2$ vs. $(1/S)(dN_\mathrm{ch}/d\eta)$ shows excellent universality over the entire collision energy range between 7.7 and 2760\,$A$\,GeV. But for MC-KLN (Fig.~\ref{fig2}b), lower collision energies result in larger $v_2/\epsilon_2$ values at the same charged hadron multiplicity density (as previously shown in \cite{Shen:2011eg}). We find that the main reason for the different collision energy dependences of the two models lies in their different behavior of the initial overlap area $S$: As the collisions become more peripheral, $S$ decreases more rapidly in the MC-KLN model than in the MC-Glauber model \cite{Shen:2012vn}. This slightly faster drop of $S$ in the MC-KLN model shifts the ``universal'' scaling curves in Fig.~\ref{fig2} to the right and shrinks the covered range in $(1/S) dN_\mathrm{ch}/d\eta$. The different $\sqrt{s}$-dependences of $v_2/\epsilon_2$ as function of $dN_\mathrm{ch}/d\eta$ in Figs.~\ref{fig2}a and \ref{fig2}b thus reflect primarily the fact that the shape of the initial profiles evolves differently with centrality in the two initialization models. 

\bigskip
\noindent
{\bf ``Saturation'' of differential elliptic flow: }
%
%=======================================
\begin{figure*}
\begin{minipage}{0.79\linewidth}
  \begin{tabular}{cc}
  \includegraphics[width=0.47\linewidth,height=0.33\linewidth]{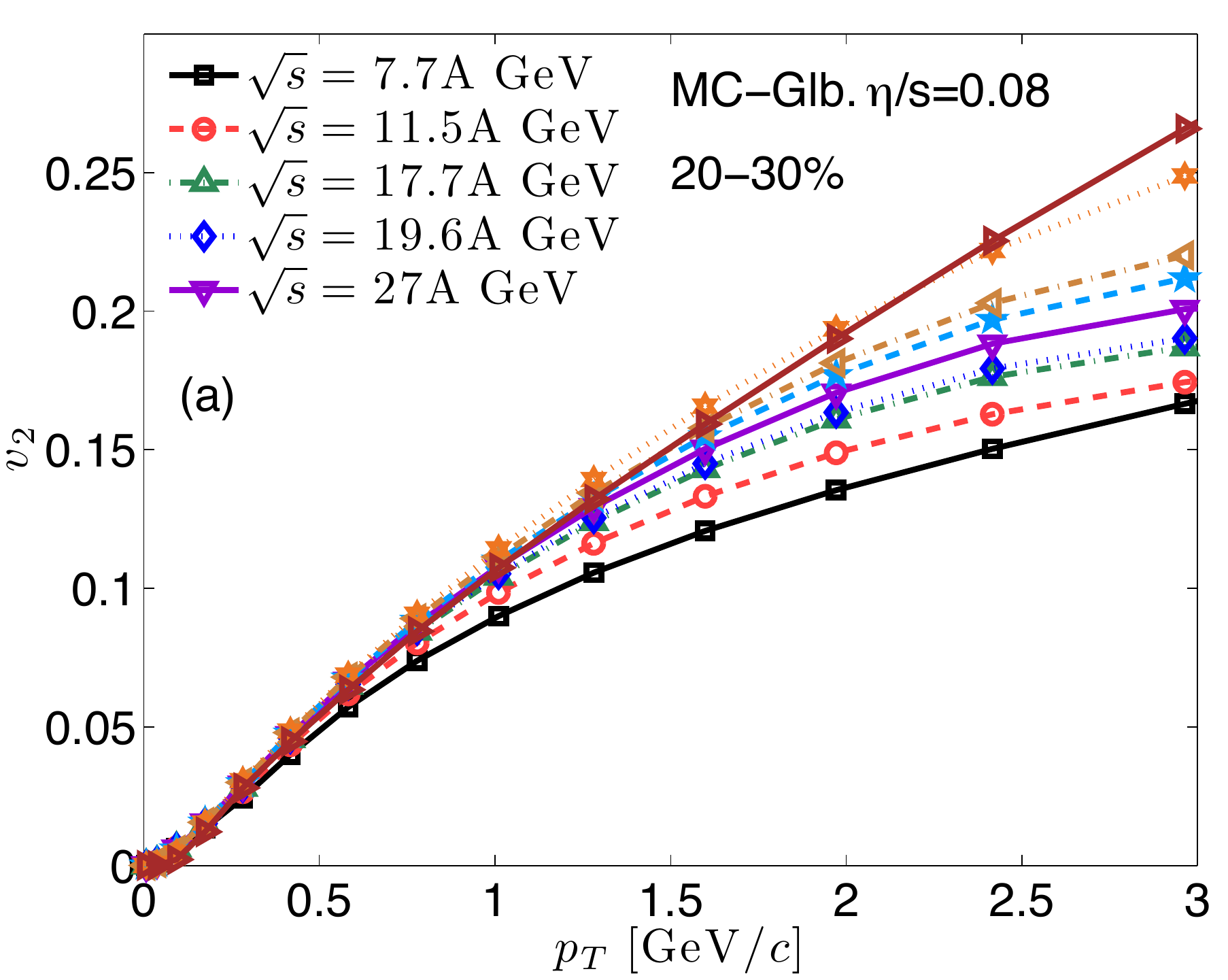} &
  \includegraphics[width=0.47\linewidth,height=0.33\linewidth]{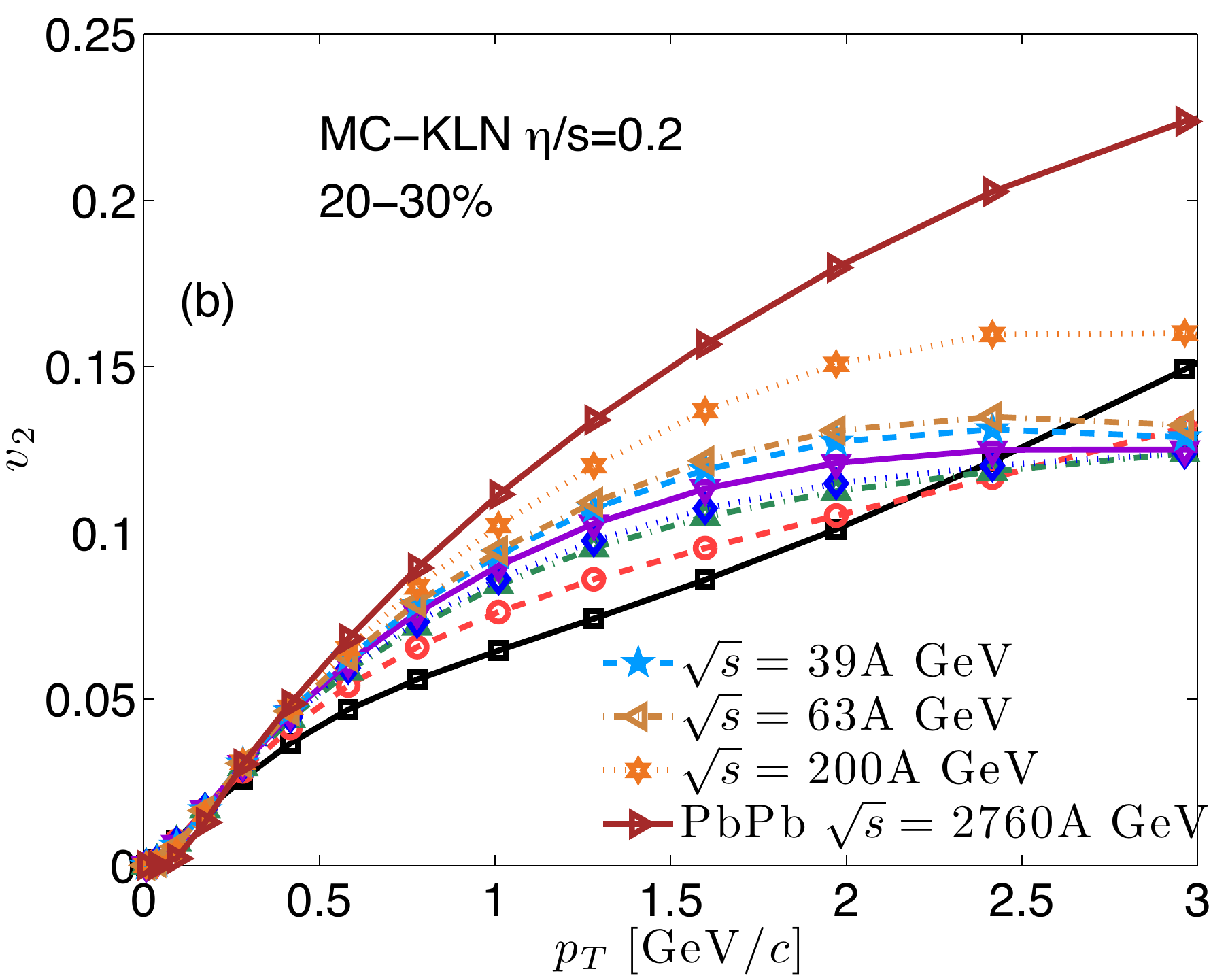} \\
  \includegraphics[width=0.47\linewidth,height=0.33\linewidth]{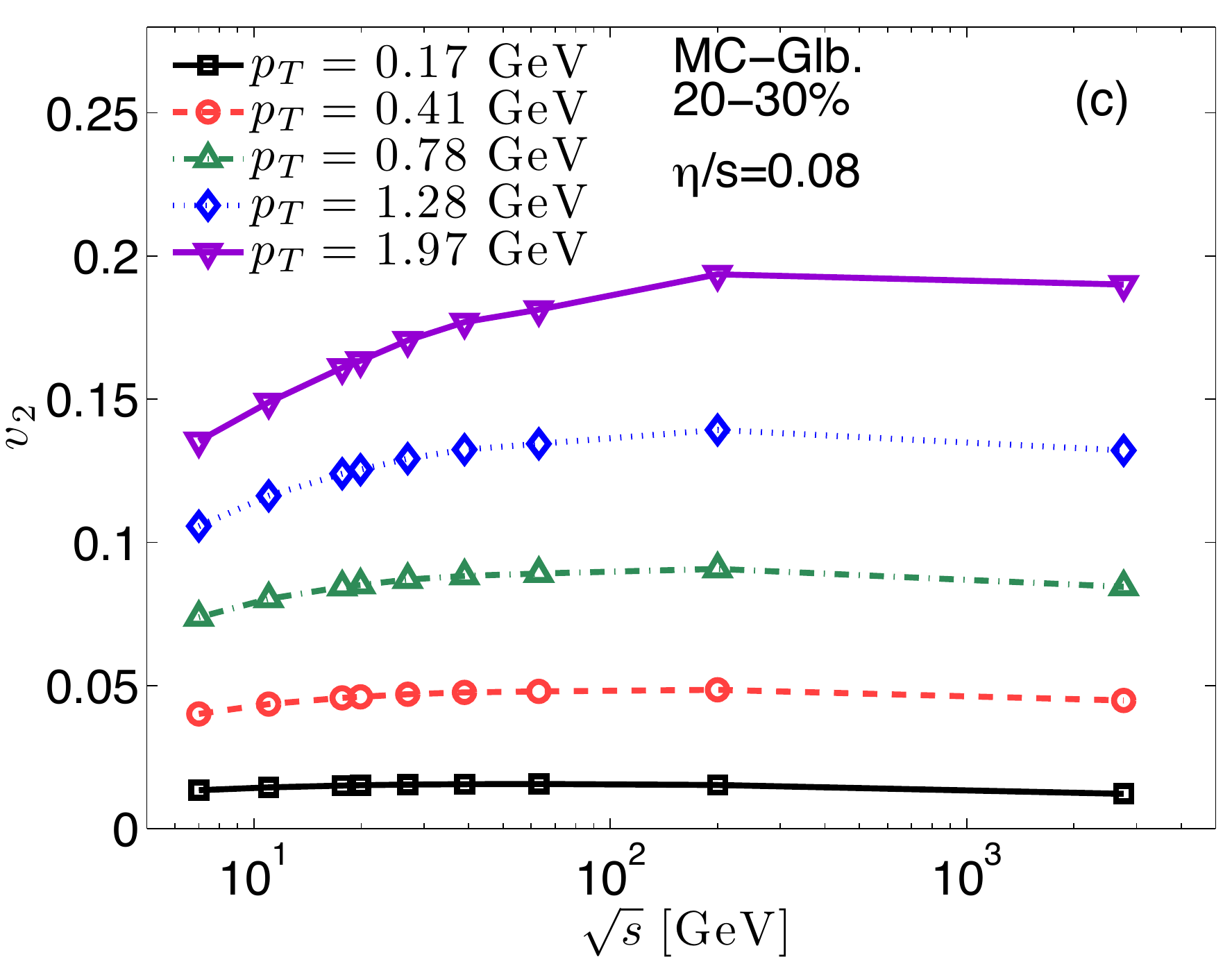} &
  \includegraphics[width=0.47\linewidth,height=0.33\linewidth]{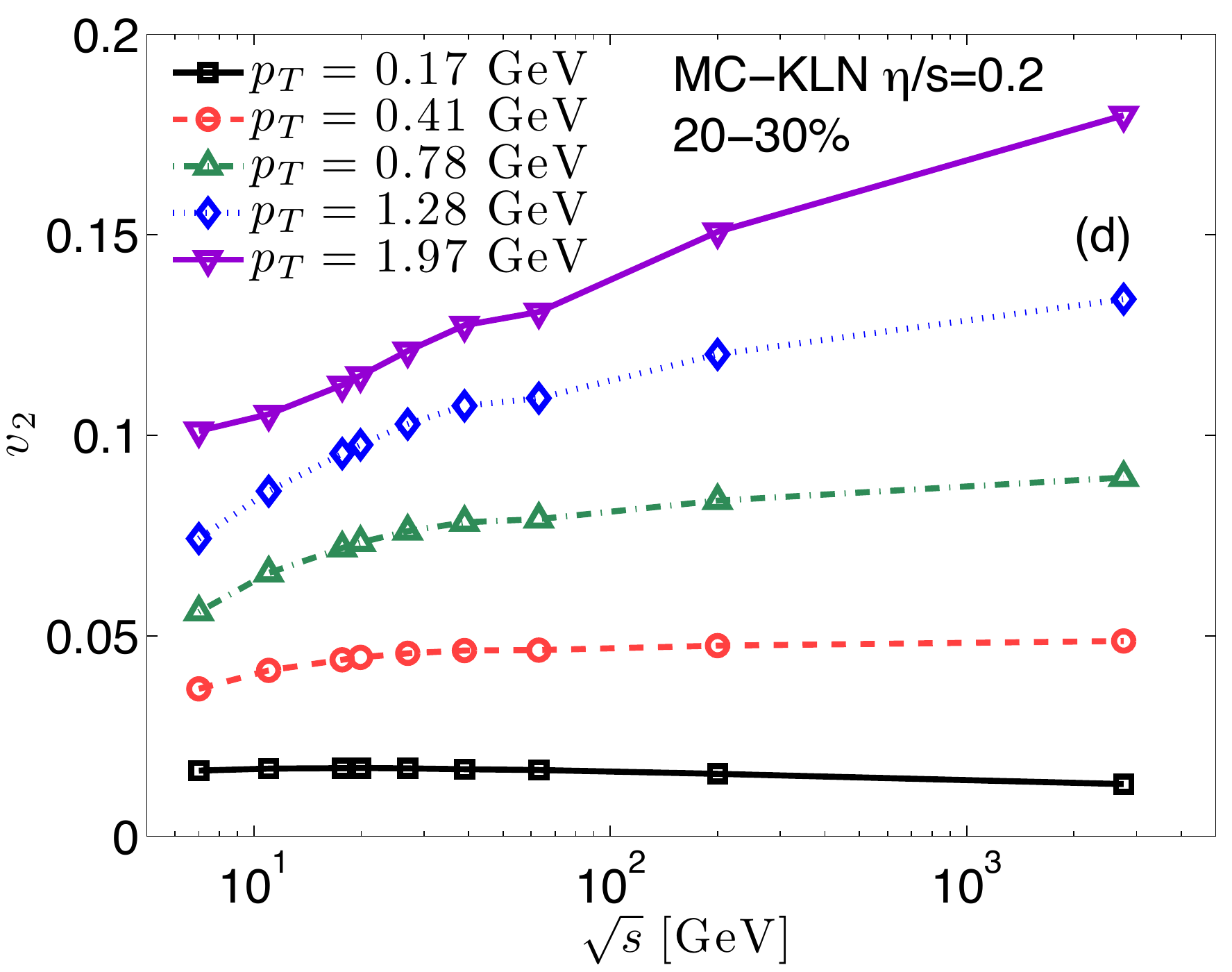}
  \end{tabular}
  \end{minipage}
  \begin{minipage}{0.2\linewidth}
  \caption{{\bf(a, b):} Differential elliptic flow of all charged hadrons at 20-30\% centrality in Au+Au and Pb+Pb collisions for different collision energies. {\bf(c, d):} $\sqrt{s}$-dependence of the differential charged hadron elliptic flow $v^\mathrm{ch}_2(p_T, \sqrt{s})$ at 5 fixed $p_T$ values below 2 GeV/$c$.}
  \label{fig3}
\end{minipage}
\end{figure*}
%=======================================
%
In Figs.\,\ref{fig3}a,b, we show the differential charged hadron elliptic flow for Au-Au or Pb-Pb collisions at 20-30\% centrality. The differential elliptic flow is affected by both total elliptic and radial flow. With MC-Glauber initial conditions the differential elliptic flow for $p_T{\,<\,}2$\,GeV/$c$ remains almost unchanged for $\sqrt{s}{\,\ge\,}39$\,$A$\,GeV. As the collision energy increases, both radial and elliptic flow increase, due to the longer fireball lifetime. Naively the increasing total elliptic flow should also lead to a larger differential $v_2$, but this tendency is counteracted by the growing radial flow which blueshifts the momentum anisotropy to larger $p_T$. For the runs with MC-KLN initial conditions we use a larger $\eta/s$ value. The resulting larger viscous effects suppress the total elliptic flow at lower collision energies more strongly than for the MC-Glauber runs, leading to a monotonous decrease of the slope of the differential $v_2(p_T)$ with decreasing collision energy. To further illustrate this point we plot in Fig.~\ref{fig3}c,d the $\sqrt{s}$-dependence of $v_2^\mathrm{ch}(p_T)$ at 5 fixed $p_T$ points. In this representation one sees that for the MC-Glauber model with $\eta/s{\,=\,}0.08$, $v_2^\mathrm{ch}$ at any fixed $p_T{\,\le\,}2$\,GeV/$c$ features, as a function of $\sqrt{s}$, a very broad maximum around top RHIC energy (200\,$A$\,GeV). For lower $p_T{\,<\,}0.5$\,GeV/$c$, heavier particles, or smaller $\eta/s$ this maximum moves towards lower $\sqrt{s}$. With the larger $\eta/s{\,=\,}0.2$ used in the MC-KLN model, the strong reduction of $v_2^\mathrm{ch}$ at low collision energies shifts the maximum of $v_2^\mathrm{ch}$ at any fixed $p_T$ towards higher $\sqrt{s}$; Fig.~\ref{fig3}d shows that for $\eta/s{\,=\,}0.2$ this observable does not peak below the top LHC energy, except for very small $p_T{\,<\,}200$\,MeV/$c$.  

\bigskip
\noindent
{\bf Freeze-out shape analysis: }
%
%=======================================
\begin{figure}
\begin{minipage}{0.6\linewidth}
\centering
  \includegraphics[width=0.7\linewidth]{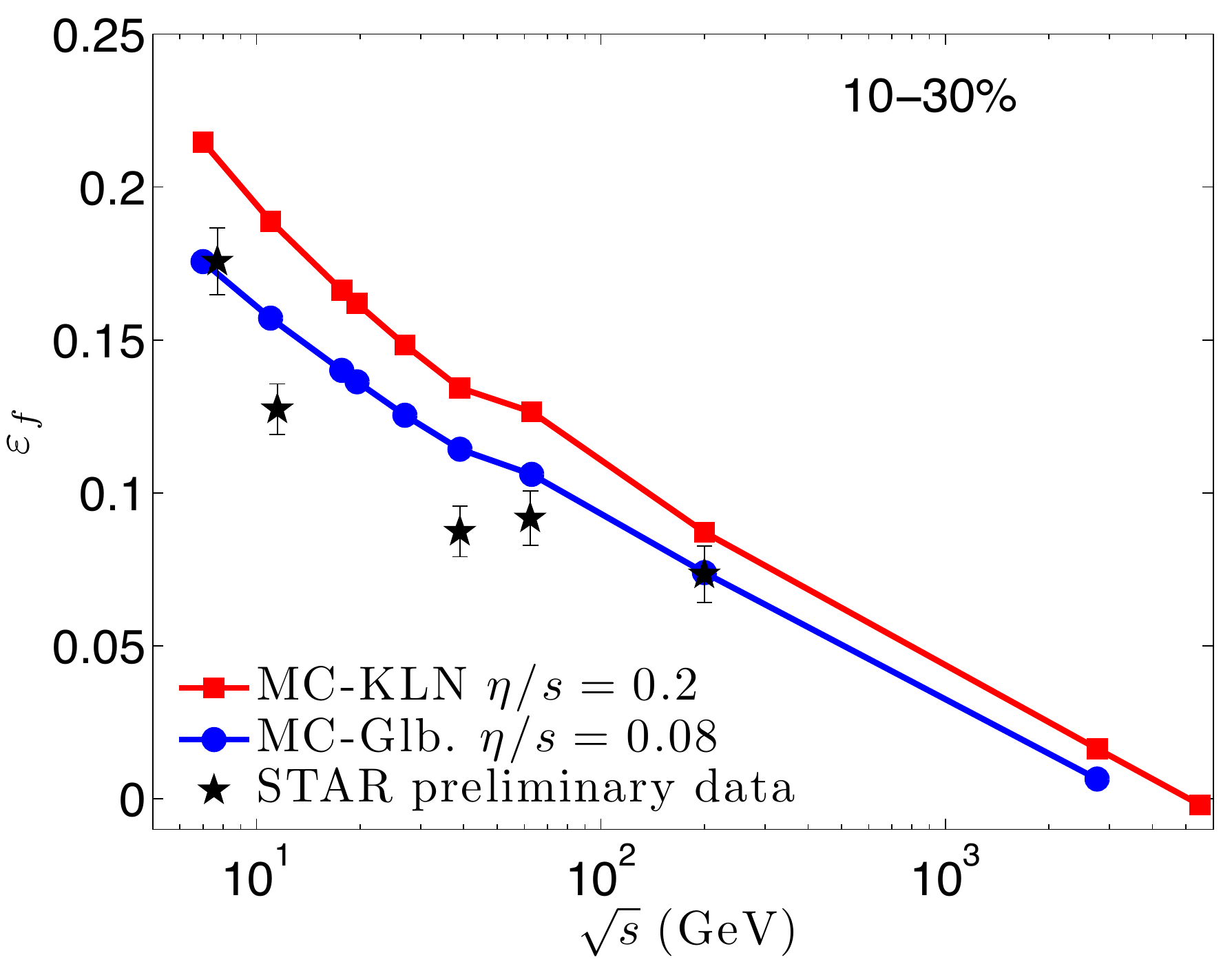}
  \end{minipage}
  \begin{minipage}{0.4\linewidth}
\caption{$\sqrt{s}$-dependence of the final spatial eccentricity $\varepsilon_\mathrm{f}$ of 
  the isothermal kinetic freeze-out surface at $T_\mathrm{dec}{\,=\,}120$\,MeV, for 10-30\% 
  centrality. The initial eccentricity is 0.26 for the MC-Glauber model and 0.32 for the MC-KLN
  model. The experimental points indicate preliminary data \cite{Anson:2011ik} from an 
  azimuthal HBT analysis by the STAR Collaboration.}
  \label{fig4}
  \end{minipage}
\end{figure}
%=======================================
% 
In Fig.~\ref{fig4}, we show the final fireball eccentricity calculated along the kinetic freeze-out surface, $T_\mathrm{dec}=120$\,MeV, as a function of collision energy.  As the collision energy increases, the final spatial eccentricity $\varepsilon_\mathrm{f}$ decreases monotonically for both MC-Glauber and MC-KLN models. This is because at higher collision energy the fireball has longer lifetime to decompress its original deformation and become more isotropic. In Fig.\,\ref{fig4}, we compare our results with recent STAR data from an azimuthal HBT analysis \cite{Anson:2011ik}. Our calculations qualitatively agree with the experimental data and reproduce the trend of the collision energy dependence of $\varepsilon_\mathrm{f}$. MC-Glauber runs with $\eta/s{\,=\,}0.08$ quantitatively reproduce the data at $\sqrt{s}{\,=\,}200\,A$\,GeV while underpredicting the final eccentricity by $\sim$10\% at lower energies. MC-KLN initial conditions with $\eta/s{\,=\,}0.2$ result in 15-20\% larger final eccentricities, due to the $\sim$20\% larger initial eccentricities of the MC-KLN profiles. Within the explored range, we found weak sensitivity of these curves to $\eta/s$. Extending our calculations to LHC energy we predict that $\varepsilon_\mathrm{f}$ will approach zero around $\sqrt{s}{\,=\,}2.76$-5.5\,$A$\,TeV. 

\bigskip
\noindent
{\bf Acknowledgments: }
We thank Christopher Anson and Mike Lisa for stimulating discussions.This work was supported by the U.S. Department of Energy under Grants No. \rm{DE-SC0004286} and (within the framework of the JET Collaboration) \rm{DE-SC0004104}.
%
%\section*{References}


\begin{thebibliography}{99}

\bibitem{Kumar:2011de}
  L.~Kumar {\it et al.} (STAR Collaboration),
  %``Results from the STAR Beam Energy Scan Program,''
  Nucl.\ Phys.\  {\bf A862-A863}, 125 (2011).
  %[arXiv:1101.4310 [nucl-ex]].
  %%CITATION = NUPHA,A862-863,125;%%
  
\bibitem{Shi:2011ad} 
  S.~Shi {\it et al.} (STAR Collaboration),
  %``The elliptic flow in Au+Au collisions at $\sqrt{s_{NN}}$ = 7.7, 11.5 and 39 GeV at STAR,''
  arXiv:1111.5385 [nucl-ex].
  %%CITATION = ARXIV:1111.5385;%%
  
\bibitem{QM12BES}
  See also related presentations by the STAR and PHENIX Collaborations in this volume.

\bibitem{Shen:2012vn} 
  C.~Shen and U.~Heinz,
  %``Collision Energy Dependence of Viscous Hydrodynamic Flow in Relativistic Heavy-Ion Collisions,''
  Phys.\ Rev.\ C {\bf 85}, 054902 (2012)
  [arXiv:1202.6620 [nucl-th]].
  %%CITATION = ARXIV:1202.6620;%%
  
\bibitem{Huovinen:2009yb}
  P.~Huovinen and P.~Petreczky,
  %``QCD Equation of State and Hadron Resonance Gas,''
  Nucl.\ Phys.\  {\bf A837}, 26 (2010).
  %[arXiv:0912.2541 [hep-ph]]. 
  
\bibitem{Shen:2010uy}
  C.~Shen, U.~Heinz, P.~Huovinen and H.~Song,
  %``Systematic parameter study of hadron spectra and elliptic flow from
  %  viscous hydrodynamic simulations of Au+Au collisions at sqrt(s_NN) =
  %  200 GeV,''
  Phys.\ Rev.\ C {\bf 82}, 054904 (2010).
  %[arXiv:1010.1856 [nucl-th]].
  %%CITATION = ARXIV:1010.1856;%%
  
\bibitem{Song:2010mg}
  H.~Song, S.~A.~Bass, U.~Heinz, T.~Hirano and C.~Shen,
  %``200 A GeV Au+Au collisions serve a nearly perfect quark-gluon liquid,''
  Phys.\ Rev.\ Lett.\ {\bf 106}, 192301 (2011);
  %%CITATION = ARXIV:1011.2783;%%
%\bibitem{Song:2011hk}
  %H.~Song, S.~A.~Bass, U.~Heinz, T.~Hirano and C.~Shen,
  %``Hadron spectra and elliptic flow for 200 A GeV Au+Au collisions from
  %viscous hydrodynamics coupled to a Boltzmann cascade,''
  and Phys.\ Rev.\ C {\bf 83}, 054910 (2011).
  %[arXiv:1101.4638].
  %%CITATION = ARXIV:1101.4638;%%
  
\bibitem{Adler:2004zn} 
  S.~S.~Adler {\it et al.}  [PHENIX Collaboration],
  %``Systematic studies of the centrality and s(NN)**(1/2) dependence of the d E(T) / d eta and d (N(ch) / d eta in heavy ion collisions at mid-rapidity,''
  Phys.\ Rev.\ C {\bf 71}, 034908 (2005)
  [Erratum-ibid.\ C {\bf 71}, 049901 (2005)]
  [nucl-ex/0409015].
  %%CITATION = NUCL-EX/0409015;%%

\bibitem{Aamodt:2010cz} 
  K.~Aamodt {\it et al.}  [ALICE Collaboration],
  %``Centrality dependence of the charged-particle multiplicity density at mid-rapidity in Pb-Pb collisions at sqrt(sNN) = 2.76 TeV,''
  Phys.\ Rev.\ Lett.\  {\bf 106}, 032301 (2011)
  [arXiv:1012.1657 [nucl-ex]].
  %%CITATION = ARXIV:1012.1657;%%
  
\bibitem{Shen:2011eg}
  C.~Shen, U.~Heinz, P.~Huovinen and H.~Song,
  %``Radial and elliptic flow in Pb+Pb collisions at the Large Hadron Collider from viscous hydrodynamic,''
  Phys.\ Rev.\ C {\bf 84}, 044903 (2011).
  %[arXiv:1105.3226 [nucl-th]].

\bibitem{Anson:2011ik} 
  C.~Anson  {\it et al.} (STAR Collaboration),
  %``Energy dependence of the freeze out eccentricity from the azimuthal dependence of HBT at STAR,''
  J.\ Phys.\ G {\bf 38}, 124148 (2011).
  %[arXiv:1107.1527 [nucl-ex]].
  %%CITATION = ARXIV:1107.1527;%%

\end{thebibliography}
\end{document}